\documentclass[preprint2]{aastex}
\usepackage{graphicx}
\usepackage{txfonts}
\usepackage{natbib}
\usepackage{longtable}

\newcommand{\gtsima}{$\; \buildrel > \over \sim \;$}
\newcommand{\simgt}{\lower.5ex\hbox{\gtsima}} 

\newcommand{\pks}{PKS~0537--441}
\newcommand{\swift}{\emph{Swift}}

\begin{document}

\title{Simultaneous \swift\ and REM monitoring of the blazar \pks\ in 2005}
\shorttitle{\pks\ in optical and X-rays}
\shortauthors{E. Pian et al.} 

\author{E.~Pian\altaffilmark{1}, 
P.~Romano\altaffilmark{2,3},
A.~Treves\altaffilmark{4},
G.~Ghisellini\altaffilmark{2},
S.~Covino\altaffilmark{2},
A.~Cucchiara\altaffilmark{5}, 
A.~Dolcini\altaffilmark{4},
G.~Tagliaferri\altaffilmark{2}, 
C.~Markwardt\altaffilmark{6},
S.~Campana\altaffilmark{2}, 
G.~Chincarini\altaffilmark{2,3},
N.~Gehrels\altaffilmark{6},
P.~Giommi\altaffilmark{7}, 
L.~Maraschi\altaffilmark{8}, 
S.D.~Vergani\altaffilmark{9,10}, 
F.M.~Zerbi\altaffilmark{2}, 
E.~Molinari\altaffilmark{2}, 
V.~Testa\altaffilmark{11}, 
G.~Tosti\altaffilmark{12}, 
F.~Vitali\altaffilmark{11}, 
L.A.~Antonelli\altaffilmark{11}, 
P.~Conconi\altaffilmark{2}, 
G.~Malaspina\altaffilmark{2}, 
L.~Nicastro\altaffilmark{13}, 
E.~Palazzi\altaffilmark{13}, 
E.J.A.~Meurs\altaffilmark{9}, 
L.~Norci\altaffilmark{10}
}

\altaffiltext{1}{INAF - Trieste Astronomical Observatory, Via G.B. Tiepolo, 
11, 34143 Trieste, Italy}
\altaffiltext{2}{INAF - Brera Astronomical Observatory, Via E. Bianchi, 46, 
23807 Merate (LC), Italy}
\altaffiltext{3}{Department of Physics, University of Milano-Bicocca,  Piazza 
delle Scienze, 3, 20126 Milan, Italy}
\altaffiltext{4}{Department of Physics and Mathematics, University of Insubria, 
Via Valleggio, 11, 22100 Como, Italy}
\altaffiltext{5}{Department of Astronomy and Astrophysics, Pennsylvania State 
University, University Park, PA 16802, USA}
\altaffiltext{6}{NASA Goddard Space Flight Center, Code 661, Greenbelt, MD 20771, 
USA}
\altaffiltext{7}{ASI Science Data Center, ASDC c/o ESRIN, Via G. Galilei, 00044
Frascati, Italy}
\altaffiltext{8}{INAF - Brera Astronomical Observatory, Via Brera, 28, 
20121 Milano, Italy}
\altaffiltext{9}{Dunsink Observatory - DIAS, Dunsink Lane, Dublin 15, Ireland}
\altaffiltext{10}{School of Physical Sciences and NCPST, Dublin City University -
Dublin 9, Ireland}
\altaffiltext{11}{INAF - Rome Astronomical Observatory, Via di Frascati, 33, 
00040 Monte Porzio Catone, Italy}
\altaffiltext{12}{Department of Physics, University of Perugia, Via A. Pascoli, 
Perugia, Italy}
\altaffiltext{13}{INAF - Institute of Space Astronomy and Cosmic Physics, 
Via Gobetti, 101, 40129 Bologna, Italy}


\begin{abstract}The blazar \pks\ has been observed  with all instruments of 
the \swift\ satellite
between the end of 2004 and November 2005. The BAT monitored it recurrently 
for a total of 2.7 Ms, and the XRT and UVOT pointed it on 
seven occasions for a total 
of 67 ks, making it one of the AGNs best monitored
by \swift.  
The  automatic optical and near-infrared
telescope REM  has monitored simultaneously  the source at  all times.
In January-February 2005 \pks\  has been detected  at its brightest  in optical and
X-rays: more than a factor of 2 brighter 
in 
X-rays and about a factor 60 brighter in the optical than observed in December 2004. 
The July 2005 observation recorded a
fainter X-ray state, albeit still brighter than the historical average.  
The simultaneous optical state, monitored  by both
\swift\ UVOT and  REM, is high, and in the VRI  bands it is comparable
to what was recorded  in  early January 2005, before the outburst.  
In  November  2005, the  source
subsided both in X-rays and optical to a quiescent state, having decreased
by factors of $\sim$4 and $\sim$60 with respect to the January-February 2005 
outburst, 
respectively.  Our
monitoring shows  an overall well correlated optical  and X-ray decay,
with no measurable time lag larger than about 1 month. 
On the shorter time scales (days or hours), there is
no obvious correlation between X-ray and optical variations, but the
former tend to be more pronounced, opposite to what is observed on
monthly time scales.  
The widely different amplitude of the long term variability in
optical and X-rays is very unusual and makes this observation a unique
case study for blazar activity.  The spectral energy distributions
are
interpreted  in  terms of the synchrotron  and  inverse Compton mechanisms
within a jet where the plasma radiates via internal shocks and the
dissipation depends on the distance of the emitting region from the 
central engine.
\end{abstract}



\section{Introduction}

Multiwavelength variability is the most effective diagnostic tool of the properties of
extragalactic jets and  of their central engines.  Due to the  orientation of
their jets  - nearly aligned  to our line  of sight - blazars  allow a
better insight into  their inner regions than other  radio loud Active
Galactic Nuclei (AGN) do, because the intrinsic flux variability
is  magnified by  relativistic  effects.  Past observations  of
blazars  have identified the  active emission  mechanisms (synchrotron
process at frequencies  up to the soft, and  occasionally hard, X-rays
and inverse Compton scattering at higher energies, Ulrich, Maraschi \&
Urry 1997; Pian et al. 1998; Tagliaferri et al. 2003; 
Krawczynski et al. 2004; Dermer \& Atoyan 2004;
B{\l}a\.zejowski et al.
2005; Sokolov \& Marscher 2005; Aharonian et 
al. 2006; Albert et al. 2006; Kato, Kusunose \& Takahara 2006; Massaro et al. 2006;
Raiteri et al. 2006). Yet, the structure
of the jet, the mechanisms of the energy transfer from the central engine
to the emitting particles, and the dissipation processes  along the
jet are not clear.  Intensive monitorings and good coverage at all
frequencies are necessary to explore the multiwavelength
variability of blazars  to its full extent  and to understand how the jet interacts with other
circumnuclear components to produce the radiation.
The \swift\ satellite (Gehrels et al.\  2004), 
with its easy and flexible scheduling, can be
optimally employed for the observation of bright blazars (Giommi et al. 2006;
Sambruna et al. 2006; Tramacere et al. 2006).

The blazar \pks\ ($z = 0.896$)  is a bright emitter at all frequencies
from radio  to gamma-rays.   In the latter  band it has  been observed
many  times by  EGRET  and  detected in  different  states (Treves  et
al. 1993;  Hartman et al.  1999;  Pian et al.  2002).   The source was
targeted  for long term optical  and NIR  monitoring with  the automatic
optical/near-IR  60cm  telescope  Rapid   Eye  Mount  (REM,  Zerbi  et
al.\ 2001; Chincarini et al. 2003; Covino et al.\ 2004)  
in December 2004 - March 2005, when the
blazar exhibited  a flare  with a 
time scale of about a month (Dolcini et
al.\ 2005).  At that time, \swift\ 
had observed \pks\ for calibration purposes. On 25 June 2005, the {\it
RXTE} All  Sky Monitor recorded  a high X-ray  (2-10 keV) state  of the
source, with a flux of ($13 \pm 5$) milliCrab.  The REM monitoring in
the  optical  also  revealed  the  blazar  to  be  active  (Covino  et
al. 2005).  Based on these  alerts, we requested observations of \pks\
as  a Target  of Opportunity  with  \swift.  A  first observation  was
scheduled in July 2005. A  second \swift\ visit took place in November
2005, in order to monitor the long-term behavior  of the source after
the June  2005 outburst.   We report here  the results of  all \swift\
observations of  \pks, and of the  simultaneous REM observations  in July
and November 2005.

\section{Data  acquisition, reduction and analysis}

\subsection{X-ray observations}

\subsubsection{\swift/BAT}

\pks\ was often in the field  of view of  the \swift\ Burst  Alert Telescope
(BAT, 15--150 keV) 
from December 2004 to November 2005. 
The BAT data  were analysed using the  standard BAT
analysis software distributed within FTOOLS v6.0.5.  
Although the blazar is not detected in individual \swift\ orbits by BAT, 
averaging the BAT signal during all periods of observation
results in significant flux detection.
A spectrum of the
integrated data  set was  extracted and fitted  to a  single power-law
$F_\nu \propto  \nu^{-\beta}$
with spectral index $\beta = 0.5  \pm 0.5$ (reduced  $\chi^2 = 1.03$).
The flux in the  15-150 keV band is $(3.2_{-2.3}^{+0.9})\times10^{-11}$
erg~cm$^{-2}$~s$^{-1}$.

\subsubsection{\swift/XRT}

The blazar was observed with the
X-Ray  Telescope   (XRT, 0.2-10 keV, Burrows et al.\ 2005) 
in December 2004-February 2005, July 2005, and November 2005.  The
monitoring is organized in seven observations, four of which were obtained
during  the \swift\  XRT  calibration  phase.  
The  XRT  data  were first  processed  by  the  \swift\  Data  Center  at
NASA/GSFC into Level 1  products (calibrated and quality-flagged event
lists).  Then  they were  further  processed  with  the latest {\tt  Heasoft}
release\footnote{http://swift.gsfc.nasa.gov/docs/software/lheasoft}
(v6.0.5) to produce  the final cleaned event lists.  In particular, we
ran the  task {\tt xrtpipeline} (v0.10.3)  applying standard filtering
and  screening criteria, i.e.,  we cut  out temporal  intervals during
which  the CCD  temperature was  higher than  $-47$ $^\circ$C,  and we
removed hot and flickering pixels which are present because the CCD is
operating  at a  temperature  higher than  the  design temperature  of
$-100$ $^\circ$C  due to a failure  in the active  cooling system.  An
on-board  event threshold  of $\sim$0.2  keV was  also applied  to the
central pixel, which has been  proven to reduce most of the background
due to  either the bright  Earth limb or  the CCD dark  current (which
depends on the  CCD temperature).  Given the low  rate of \pks\ during
the observing  campaign ($<0.5$  counts s$^{-1}$ in  the $0.2-10$\,keV
range), 
we only considered photon  counting data for our  analysis (PC; see
Hill et al. 2004 for a full  description of read-out modes) and
further selected  XRT grades 0--12, (according  to \swift\ nomenclature;
Burrows et al. 2005).   A  summary  of the  XRT
observations is reported in Table~1.  

\pks\ was  detected in the XRT data at  the coordinates  RA(J2000)$=05^{\rm h}
38^{\rm  m}  50\fs38$,  Dec(J2000)$=-44^{\circ} 05\arcmin  09\farcs1$,
with  an estimated  uncertainty  of 3\farcs5  arcseconds radius  (90\%
containment).
This position takes into account the correction for the
misalignment  between the  telescope  and the  satellite optical  axis
(Moretti et al. 2006), and is consistent with the source catalog position 
(ICRS coordinates are RA(J2000)$=05^{\rm h}
38^{\rm  m}  50\fs36$,  Dec(J2000)$=-44^{\circ} 05\arcmin  08\farcs94$).

We extracted the source events in a circle with a radius of 30 pixels
($\sim71$\arcsec), which  corresponds to $\sim  94$\% of the  XRT PSF.
To account for the background,  the data were also extracted within an
annular region (radii 55 and 95 pixels) centered on the source and 
devoid of background sources.  

The  source  and background  spectra  were  extracted  in the  regions
described  above.  Ancillary  response files  were generated  with the
task {\tt xrtmkarf}, and  account for differences in extraction regions and
PSF corrections.  We used  the latest spectral redistribution matrices
(RMF,  v008).   The adopted  energy  range  for  spectral fitting  is
0.3--10 keV,  and all data were  rebinned with a minimum  of 20 counts
per energy bin to allow  $\chi^2$ fitting within XSPEC (v11.3.2).  The
only  exception was the observation of 23  December 2004,  when the
number of  counts was limited ($\sim140$)  and Cash (1979)
statistics  was  appropriate,  therefore  ungrouped  data  were  used
instead.  

The spectra do not  exhibit significant features, either in absorption or
emission, superimposed on the power-law continuum (see a representative
spectrum in Figure~1).
We considered an  absorbed power-law model, with the neutral
hydrogen  column  kept  fixed  to  its Galactic  value  ($2.91  \times
10^{20}$ cm$^{-2}$, Murphy et al.  1996), and the spectral index left as
a free parameter.  The fit results are reported in Table~1.

\subsection{Optical observations}

The  \swift\ UltraViolet-Optical  Telescope (UVOT, Roming et al.\ 2005)  
observed  \pks\ in
July  (only  U,B  and  V  filters) and  November  2005  (all  filters)
simultaneously with the XRT.  The  log of the observations is reported
in Table~2. 
UVOT data were taken also in December 2004 and  January 2005, 
but due to early orbit 
checkout and calibration, they are not reliable, and therefore
we have not used them.
The  data analysis  was performed using  the ``uvotsource''
task    included   in    the   latest  {\tt  Heasoft}   software (see Section 
2.1.2).  We
subtracted the background, and corrected for the coincidence loss effect
(similar to  the pileup for the XRT)  in the case of  a bright source.
The magnitudes  were converted into fluxes using  the latest in-flight
flux calibration factors and zero-points.

REM acquired photometry of the  AGN from December 2004 to March 2005,
and also  in July and in  November 2005, with  various combinations of
filters.  The data reduction followed standard procedures (see Dolcini
et al. 2005).   The log of the July and  November 2005 observations is
reported  in   Table~2,  while  the   details  of  the   previous  REM
observations have been presented in Dolcini et al. (2005).

\section{Results}

\subsection{Multiwavelength light curves}

With a total \swift/XRT exposure of 67 ks, \pks\ is one of the blazars 
best monitored by this instrument. 
Figure~2  shows the  observed (i.e.,  
not  corrected for
Galactic absorption), background-subtracted  light curves extracted in
the 0.2--1 keV and 1-10  keV energy bands.  For direct comparison with
the X-rays, 
the merged UVOT and REM light
curve  in  the  V  band (covered by both instruments) 
is also shown 
in Figure 2.
The UVOT V-band  fluxes have been reduced to the
central wavelength of the  REM V-band observations using the power-law
$F_\nu \propto  \nu^{-\beta}$ which best fits the  optical spectrum in
July ($\beta = 1.84 \pm 0.04$) and November 2005 ($\beta = 1.26 \pm 
0.05$).  The full transformation equation is 
$F_{UVOT,5505 {\rm \AA}} = (5505/5460)^{\beta} \times F_{0,UVOT-V} \times
10^{-0.4 V_{UVOT}}$, where $F_{0,UVOT-V}$ is the flux corresponding to
zero UVOT V magnitude (equal to $3.19 \times 10^{-9}$ and 
$3.17 \times 10^{-9}$ 
erg~s$^{-1}$~cm$^{-2}$~\AA$^{-1}$ for the July and November 2005 
measurements, respectively).
The REM and
UVOT V-band data taken closest in time (12 July 2005) differ by 
$\sim$13\%, the REM flux being lower than the UVOT flux. This difference
may be intrinsic,  since the REM and UVOT  observations are about
6-7 hours apart, however, it  is within the sum of the 
statistical uncertainties (see Table~2) and the systematic
errors due to flux transformation and calibration of the two instruments
(estimated to be no less than $\sim$5\% altogether).

The  optical  V-band and  X-ray light  curves are  highly correlated,
however,  the V-band  flux
varies  with much  higher amplitude.   To  the initial  factor of  60
optical variation detected  with REM between end of  December 2004 and
early February  2005 -- noted  and discussed in  Dolcini et  al. (2005) --
corresponds a variation  of only a factor $\sim$2  of both soft (0.2-1
keV) and hard  (1-10 keV) X-ray flux (Fig.~2).  Thereafter,  the flux decays
nearly monotonically up  to November 2005 both in  optical and X-rays, with
overall amplitudes of factors of $\sim$60 and $\sim$4, respectively.
The variability indices of the X-ray light curves, defined as the ratios 
between the flux standard deviation around the mean flux  and the mean flux
itself ($\sigma/<f>$), are 0.375 and 0.423 for the hard and soft X-rays, 
respectively, consistently lower than the optical variability index, 1.434 (the variability
indices have been computed from the original datasets, i.e. before applying 
the temporal binning adopted in Figure 2).
The time  behavior of the hardness  ratio between the  bands 0.2-1 keV
and 1-10 keV shows no clear long term trend: the spectrum hardens
up to summer 2005 and softens thereafter, but only with marginal significance 
(Fig.~3a).

In Figure~4 
portions  of the light curves are reported  in smaller time
intervals.  The X-ray flux presents  an almost fully resolved flare on
27-28 January  2005 with somewhat  higher total amplitude in  the soft
than in the hard band (factors of $\sim$2 and $\sim$1.5, respectively,
see Fig.~3b and Fig.~4a).   The correlated optical and X-ray behavior
on short  time scales (days to hours) has no precise character: 
while the limited simultaneous X-ray  and optical sampling
in July  2005 shows a well correlated  decay in the 2  bands, with the
X-ray flux declining faster than the  optical flux (Fig.~4b), the factor of
2 X-ray variations in November 2005 have no counterpart in UV-optical,
where flickering of at most $\sim$10\% is observed (Fig.~4c,d).

Obviously, 
the better long term sampling available at optical wavelengths favours the
detection of day time scale variations in optical with respect to the X-rays.
This is relevant when attemping to determine a possible time lag between the optical and
X-ray light curves.
While we can 
constrain the occurrence of the optical maximum (formally observed on 5 February 2005) 
within the time window 3-12 February 2005, the X-ray light curve 
maximum is much less well constrained.
Fig. 4a indicates that the X-ray observed maximum occurred between 27 and 28 January 2005.  
If this is the absolute peak of the X-ray light curve, and it is correlated with that
in the optical, then it has preceded the optical maximum by at least one week in the
observer frame. 
However, given the sampling of the X-ray light curve,
we cannot  
exclude
more intense and unobserved 
flares 
preceding or following the observed X-ray maximum by time intervals of up to 
$\sim$1  month.  
Therefore, this is our upper limit on the
time lag of the
correlation between the X-ray and optical light curves.

\subsection{Broad-band spectrum}

In  Figure~5 we  report the  broad-band spectral  energy distributions of  
\pks\ at three epochs during  our \swift\ and  REM campaign,
representative of three different emission states: 24-25 February, 12
July and 24 November 2005 for the bright, intermediate and low state,
respectively.  The spectral energy distribution of 24 November 2005 has been selected because the
\swift/UVOT observations made on that day 
cover the near-UV wavelengths (1930-2600 \AA), 
unlike those of
17 November, that are limited to the UBV filters (Table~2).  However, no
strictly simultaneous REM data are available on 24 November.  The REM data
of 20 and 30 November 2005 have been used instead, and interpolated at the date
of 24 November.  We exclude that possible variability between 20 and 30 November
2005 may significantly affect the reliability of the REM fluxes obtained through 
interpolation: no large variability is observed in this period (see Table~2); the 
UVOT and REM V-band points are consistent (see Fig.~5); we have verified that 
the shape of the 
near-IR-to-near-UV spectrum of 24 November 2005 is similar to that of 17 November
2005 (constructed with data simultaneous within 1 day), 
in the common wavelength range (3400--16000 \AA).

Whenever more
than one  UVOT or REM  measurement is available  at a given  date and
filter,  we  take the  flux  average.   The  associated error  is  the
standard deviation when three or  more data points are averaged. When
only two measurements  are available, the error is  the larger of the
two individual errors, or the flux difference, whichever is larger.
The  X-ray data  are  corrected for  photoelectric  absorption by  the
Galactic  neutral hydrogen  as  described in  Section  2.1.2, and  the
near-infrared  to ultraviolet  data  are corrected  for Galactic  dust
absorption  with $E(B-V)  =  0.037$ (Schlegel,  Finkbeiner, \&  Davis
1998),  using  the extinction  law  of  Cardelli,  Clayton, \&  Mathis
(1989).
For comparison,  we have reported also  the historical multiwavelength
spectra  obtained  in  1991-1992 and  1995  (see  Pian  et al.  2002)
and  the
non-simultaneous IRAS,  ISO, HST and BeppoSAX data taken  at various epochs
(Padovani et al.   2006; Pian et al. 2002).

The  2005   optical  spectra,  spanning   a  factor  of   $\sim$50  in
normalization, bracket the  historical optical observations.  They are
described by single  power-laws and are steeper at  higher states. The
near-IR  flux varies  with lower  amplitude.  At  the lowest  state of
November 2005,  we note  a large discrepancy  between the  optical and
infrared fluxes:  the H-band flux exceeds  by a factor  of $\sim$4 the
extrapolation of the optical spectrum to the H-band wavelengths.
The  flatness of  the November  optical-UV spectrum  and  the spectral
discontinuity between  the optical and near-IR  wavebands suggest that
in  the lower states different emission components play a role in shaping
the spectrum.  In particular, radiation produced  by the accretion  disk
may partially account for the optical-UV spectrum.
This behavior  is reminiscent of that seen in 3C~279
(Pian et al. 1999).

The X-ray fluxes detected by XRT  encompass
both the BeppoSAX and  ROSAT states.  The
steadiness of the XRT spectral slope over time, as opposed to a factor
4 variation  in the normalization, is remarkable.   The X-ray spectral
shape is  also very  similar to that  of the BeppoSAX  spectrum, which
covers an energy range similar to that covered by XRT.

The BAT spectrum is dominated by the episodes of more intense activity 
of January-February and July 2005, and despite its large  uncertainty gives a
good estimate of the spectral shape at hard X-rays in high state.

\section{Discussion}

The character of the multiwavelength variability detected by our monitoring is extremely unusual:
the outburst of \pks, jointly monitored by XRT and REM from its rise 
in December 2004 - January 2005
to its long decay ended in November 2005, has a remarkably higher 
amplitude at optical (factor of $\sim$60) than at  X-ray frequencies  
(factor of $\sim$4).   These obviously represent only lower limits to the 
variability, the intrinsic amplitude of which may be contaminated by a 
constant component that is more relevant in X-rays than in the optical.  
The optical spectra suggest the presence of an underlying
thermal optical component in low state (Section 3.2), presumably only 
modestly variable.  However,  trying to assess whether - and  how significantly -  
this dilutes the intrinsic multiwavelength
variability of the non-thermal flux is prone to many uncertainties.
The sampling of our monitoring indicates that the long--term decay is 
monotonic (Fig. 2), but small flares are present on day time scales (Fig. 4).

The fact that the optical--UV flux variability has a much larger 
amplitude than the simultaneous X--ray flux variability, may at first sight
be surprising. 
In a simple synchrotron self--Compton scenario, where the optical
emission is due to synchrotron radiation and the X-rays are due to
inverse Compton scattering off the synchrotron photons, one would
expect to observe the opposite if the changing parameter is the density 
of the emitting particles. 
In fact, the self--Compton emissivity scales with the square of the particle 
density, while the synchrotron emissivity varies linearly with it.
If instead the varying parameter is the magnetic field, we expect that both 
the synchrotron and the synchrotron self-Compton fluxes vary with the same amplitude.

On the other hand, in models producing the high energy emission by 
upscattering of radiation produced outside the jet ("external" Compton)
both the synchrotron and the inverse Compton fluxes vary linearly with the 
particle density.
In these models, a variation of the magnetic field could produce a 
variation of the synchrotron flux leaving almost unchanged the inverse Compton  flux.

Note also  that the X-ray and optical 
spectra in these models  derive from very different portions of the
relativistic electron distribution:  the optical emission originates  from
electrons above the spectral break, while the X-rays are produced  via inverse Compton
scattering of synchrotron or external  photons by electrons of much lower energies.

We  applied a simple, one--zone, homogeneous 
synchrotron self-Compton plus external  inverse Compton model to the different states
of PKS 0537--441.
The model is described in Ghisellini, Celotti \& Costamante
(2002). The general assumptions are the following:

\begin{itemize}

\item
The source is a cylinder of cross sectional radius
$R=\psi z$, where $z$ is the distance from the apex of the jet,
assumed to be a cone of semi-aperture angle $\psi$.
The width $\Delta R^\prime$, as measured in the comoving frame, is
assumed to be equal to $R$;

\item 
the magnetic field $B$ is homogeneous and tangled;

\item 
the blob moves with a bulk Lorentz factor $\Gamma$ 
and the viewing angle is $\theta$ (we assumed here $\theta=3^\circ$);

\item
the external radiation is produced at a fixed radius,
which can be identified with the radius of the broad line region.
We assume that 10\% of the disk luminosity is reprocessed by the
broad line region.
These assumptions should be taken with care, because, beside
the contribution of the broad line region, other processes can contribute to 
the external radiation (scattering by ionized intercloud plasma,
synchrotron radiation ``mirrored'' by the clouds and/or the walls
of the jet, reprocessing by a molecular torus; 
Sikora, Begelman, \& Rees 1994; 
B{\l}a\.zejowski et al. 2000; 
Ghisellini \& Madau 1996;
and finally, direct radiation from 
the accretion disk, Dermer \& Schlickeiser 1993; see also Celotti, Ghisellini
\& Fabian 2007);

\item
the particle distribution is the result of injection and cooling.
We calculate the random Lorentz factor $\gamma_{\rm cool}$ at which
the particles cool in one light crossing time.
If the particles are injected between $\gamma_{\rm inj}$ and $\gamma_{\rm max}$
with a power law distribution of slope $s$, 
in the ``fast cooling'' regime
($\gamma_{\rm cool}<\gamma_{\rm inj}$), we have an emitting particle distribution
$N(\gamma) \propto \gamma^{-p}$ between $\gamma_{\rm inj}$ and $\gamma_{\rm max}$
(where $p=s+1$ is the injection slope increased by one unit),
and 
$N(\gamma) \propto \gamma^{-2}$ between $\gamma_{\rm cool}$ and 
$\gamma_{\rm inj}$ (Ghisellini et al. 2002; Ghisellini et al. 1998; Ghisellini 1989).
In the models presented here, this is always the case, since  
the adopted parameters (see Table~3) imply that  the radiative 
cooling (synchrotron, synchrotron self-Compton and external Compton)
is fast, and guarantee that $\gamma_{cool}$ is always smaller 
than $\gamma_{inj}$ and close to unity (after a light crossing time).

\end{itemize}

Based on the above assumptions, the modelling of the spectral energy 
distribution yields the results shown
in Fig.~\ref{sedbat}, where we report model curves for three states of the 
source during our 2005 campaign.
We have also modeled under similar assumptions previous multiwavelength
energy distributions of \pks, presented in Pian et al. (2002), including data in the 
MeV--GeV domain from CGRO-EGRET (Fig.~\ref{sedold}).  The radiation processes at work are
synchrotron at radio-to-UV frequencies, and inverse Compton scattering 
off both synchrotron photons (self-Compton) and  external photons, 
dominating at X-ray and gamma-ray frequencies, respectively. 
The input parameters used for these models are listed in 
the upper part of Table~3. 
In the lower part we report some interesting output parameters,
namely the Doppler factor $\delta$ and the power carried 
by the jet in the form of magnetic field ($L_B$), cold protons
($L_{\rm p})$, relativistic electrons ($L_{\rm e}$), 
and produced radiation ($L_{\rm rad}$), defined as:
\begin{eqnarray}
L_B\,         &=& \,  \pi R^2 \Gamma^2 c \, U_B
\nonumber \\ 
L_{\rm p}\,   &=& \,  \pi R^2 \Gamma^2 c \int N(\gamma) m_{\rm p} c^2 d\gamma
\nonumber \\
L_{\rm e} \,  &=& \,  \pi R^2 \Gamma^2 c \int N(\gamma) \gamma m_{\rm e} c^2 d\gamma
\nonumber \\
L_{\rm rad} \, &=& \, \pi R^2 \Gamma^2 c \, U^\prime_{\rm rad} 
\end{eqnarray}
where $U_B$ and $U^\prime_{\rm rad}$ are the magnetic and
the radiation energy density measured in the comoving frame, respectively.


Usually, when applying a single--zone synchrotron and inverse Compton model,
the choice of the input parameters is not unique, especially when the
peak of the inverse Compton component is not observed, as in our case.
However, we were guided in our choice by the knowledge of the total
luminosity of the broad lines ($\sim 5\times 10^{44}$ erg s$^{-1}$, 
Pian, Falomo, \& Treves 2005), 
and the requirement that the spectra observed
here in the low optical--UV state are unusually flat because they are 
``contaminated" by the thermal accretion disk component.
We can then infer the luminosity of this component (which we have fixed to
$1.8\times 10^{46}$ erg s$^{-1}$).
Note also that in the low state of November 2005 the flux in 
the $H$ filter suggests that the synchrotron spectrum on these 
occasions is very steep.
The other requirements we have applied, which help us in choosing the
input parameters, include minimizing the total power budget and
describing the different states of the source
with a minimal change of the power carried by the jet.
The latter point is crucial, because it allows a direct test once
high energy observations -- as will be performed by GLAST -- will be
available.
In fact it is possible to obtain reasonably good fits by allowing
the jet power to vary by a large amount from state to state, being
larger in high states. 
This would however correspond to very different spectra and fluxes in the
MeV--GeV band.

Remarkably, the chosen parameters correspond closely 
to the expectations of the jet radiation model proposed by 
Katarzy\'nski \& Ghisellini (2007): dramatic variations in specific frequency
bands can be produced by relativistic jets carrying the same
amount of energy in bulk relativistic and Poynting flux form.
In fact, if blobs having the same bulk kinetic energy $\Gamma M c^2$
(where $M$ is the mass of the blob) dissipate a fraction of
their kinetic energy at different locations along the jet, 
and if the distance of the emitting region from the origin of the jet 
is directly related  to $\Gamma$, then
slow blobs will dissipate closer to the jet apex, when the blob 
is more compact, and embedded in a larger magnetic field.
In this case the external Compton scattering is reduced because the
external radiation energy density as seen in the comoving frame
($\propto \Gamma^2$) is lower, while the synchrotron radiation
($\propto B^2$) is likely to be enhanced, since we expect larger 
values of the magnetic field closer to the apex of the jet.

In this picture the key ingredient is the link between the
dissipation site and the bulk Lorentz factor at that location:
smaller Lorentz factors are required 
for smaller distances between the jet apex and the
dissipation site.
There are two scenarios: the blob could be still accelerating 
when it dissipates, or else the dissipation is the 
result of internal shocks.
In Katarzy\'nski \& Ghisellini (2007) the second scenario is adopted,
because it provides the scalings
needed to characterize completely the model.
In this scenario (see its specific application to blazars by
Ghisellini 1999, Spada et al. 2001; Guetta et al. 2004), 
faster blobs can catch up with slower ones
at a distance $z=\Gamma^2\Delta z_0$ from the jet origin, 
where $z_0$ is the initial separation of the two blobs, and $\Delta$ is their thickness.

Furthermore, in the Katarzy\'nski \& Ghisellini (2007) model,
it is assumed that the blobs
always carry the same amount of bulk kinetic energy ($\Gamma Mc^2$ is the same)
and magnetic energy ($\Gamma B^2 V^\prime$ is the same, where
$B$ and $V^\prime $ are the magnetic field and volume measured in the
comoving frame of the source, respectively).
As a result of the dissipation process, the fraction of the
available energy transferred to the emitting electrons is the same 
(i.e., the efficiency is the same).
Based on these assumptions, we can assign for all the input parameters
their scalings with $\Gamma$: therefore, when fitting two or more 
different states of the same source with this model, 
once we have chosen the parameters for one state we are left with only 
one free parameter, i.e. the bulk Lorentz factor $\Gamma$.
We can relax this by allowing the particle distribution slopes to
be changed, as well as the fraction of the electron population which is 
accelerated to relativistic energies.
This does not violate any strong requirement.

In Fig. \ref{sequence} it is shown how the choice of the present input
parameters compares with the Katarzy\'nski \& Ghisellini (2007)
prescriptions, namely 
$B\propto \Gamma^{-7/2}$, $R\propto \Gamma^2$ and 
$L_{\rm inj}^\prime \propto \Gamma^{-3}$.
We also show (bottom panel) that the power carried by the jet 
is almost constant.
The good agreement leads us to conclude that the variations
seen in this source are probably due to (small) variations in
the bulk Lorentz factor, which induces dissipation to occur at
different locations along the jet. 
Consequently,  the emitting
regions have different radii, particle densities and magnetic fields.
Note that variations of $\Gamma$ between 10 and 15 can explain
the entire observed variability.
In this respect, PKS 0537--441 is very similar to 
3C~454.3 during its large 2005 multiwavelength flare 
(Fuhrmann et al. 2006; Pian et al. 2006; Giommi et al. 2006).
For that source, the 100--fold amplitude of the optical variability 
could be accounted for by changing $\Gamma$ by a factor of less than 2 
(Katarzy\'nski \& Ghisellini 2007).

The knowledge of the spectrum of \pks\ at MeV-GeV energies is
crucial for fully constraining the models.  Therefore, 
\pks\ qualifies as a prime candidate for further
monitoring with \swift\ and for simultaneous
observations with INTEGRAL, AGILE and GLAST.

\begin{acknowledgements}

We thank P. Roming and S. Holland for assistance with the UVOT data, C. Pagani for
help with the \swift\ scheduling, and the whole \swift\ team  for
their support of the mission.   
We acknowledge use of the  quick-look results provided by the {\it RXTE} All Sky 
Monitor team.   This research  has made use of the Simbad database.  
This  work is supported  at OABr  by ASI grant
I/R/039/04 and at OATs by the contracts ASI INTEGRAL
I/R/046/04 and ASI-INAF I/023/05/0.  

\end{acknowledgements}

\clearpage


\begin{figure*}
\resizebox{\hsize}{!}{\includegraphics[angle=270]{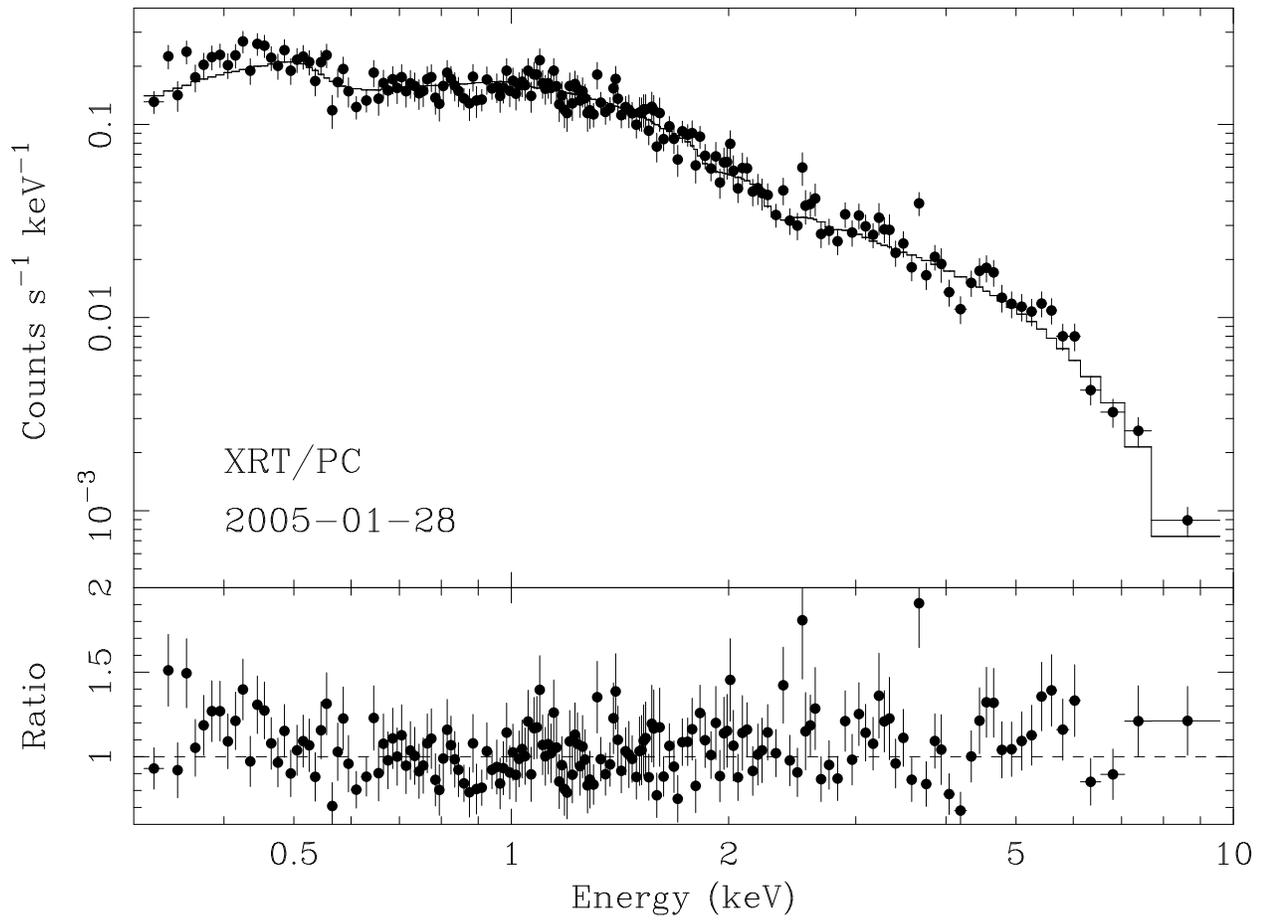}}
\caption{Observed \swift/XRT spectrum taken on 28 January 2005.
The solid stepped curve represents the single absorbed power-law
which best fits the spectrum (see  Table~1 for
spectral parameters).}
\end{figure*}


\begin{figure*}
\vskip -1.5 true cm
\resizebox{\hsize}{!}{\includegraphics[angle=0]{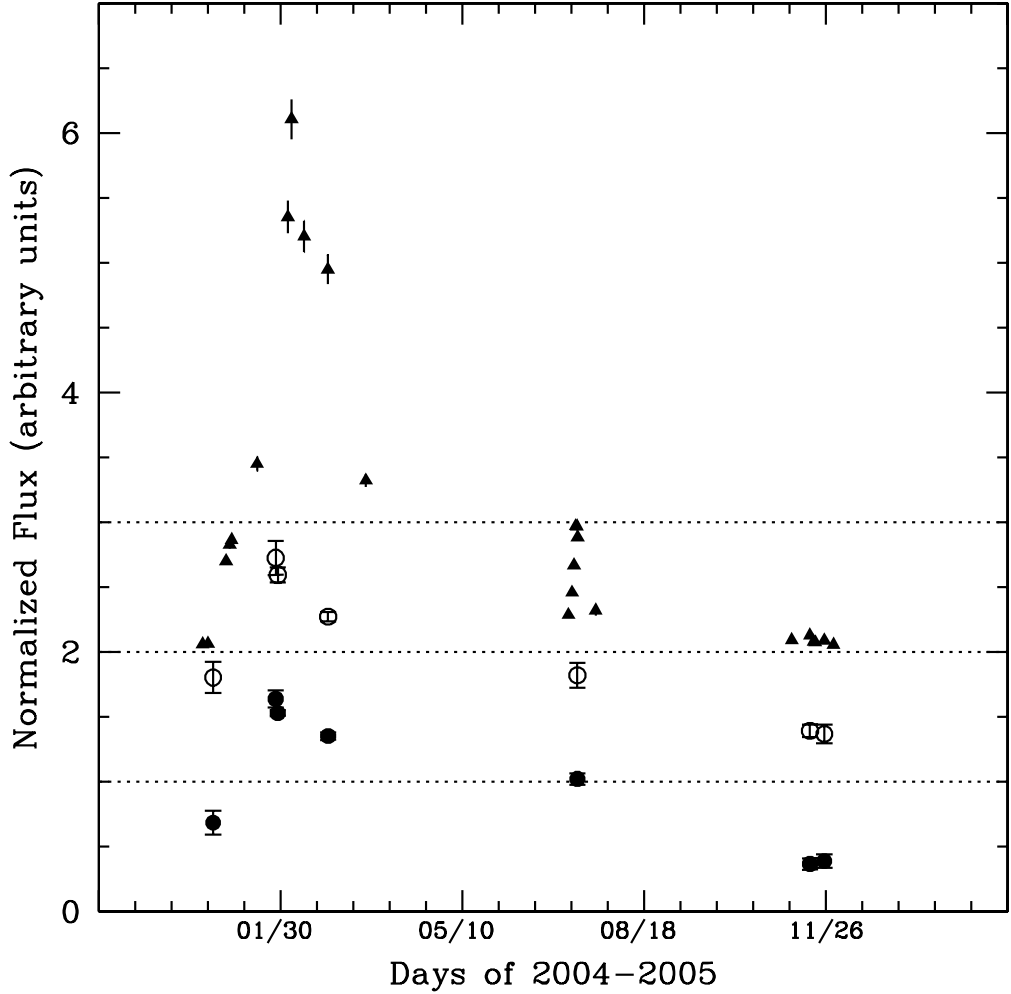}}
\vskip -1.5 true cm
\caption{\swift/XRT  background-subtracted light  curves in the  
1-10 keV (filled circles) and in the 0.2--1\,keV (open circles) 
energy bands, and optical
light  curve (triangles),  obtained from  the  merging of  the UVOT  V
filter and REM V filter observations. The signal has been averaged 
within the \swift\ pointings for the X-ray data and with a time resolution
of 1 day for the optical data. The curves are not corrected for
Galactic extinction,  and are normalized to  their respective averages
(0.136 cts~s$^{-1}$  in the 1-10  keV band, 0.084 cts~s$^{-1}$  in the
0.2-1 keV band, 6.58 mJy in  the optical band), computed on the time-binned
datasets.  The dotted horizontal lines indicate the
average values of the three light curves: for clarity, the 0.2-1
keV  band and  V-band light  curves have  been scaled  up  by additive
constants 1 and 2, respectively.  Note that this upscaling implies that the
flux ratios derived by direct inspection of the soft X-ray (0.2-1 keV) and
optical light curves
do not correspond to the real 
ones, the fluxes having been increased by constants 1 and 2, respectively. 
The maximum amplitudes of variability in optical and X-rays are a factor of $\sim$4 and $\sim$60, respectively.}
\label{pks:mwlcs}
\end{figure*}


\begin{figure*}
\resizebox{\hsize}{!}{\includegraphics[angle=0]{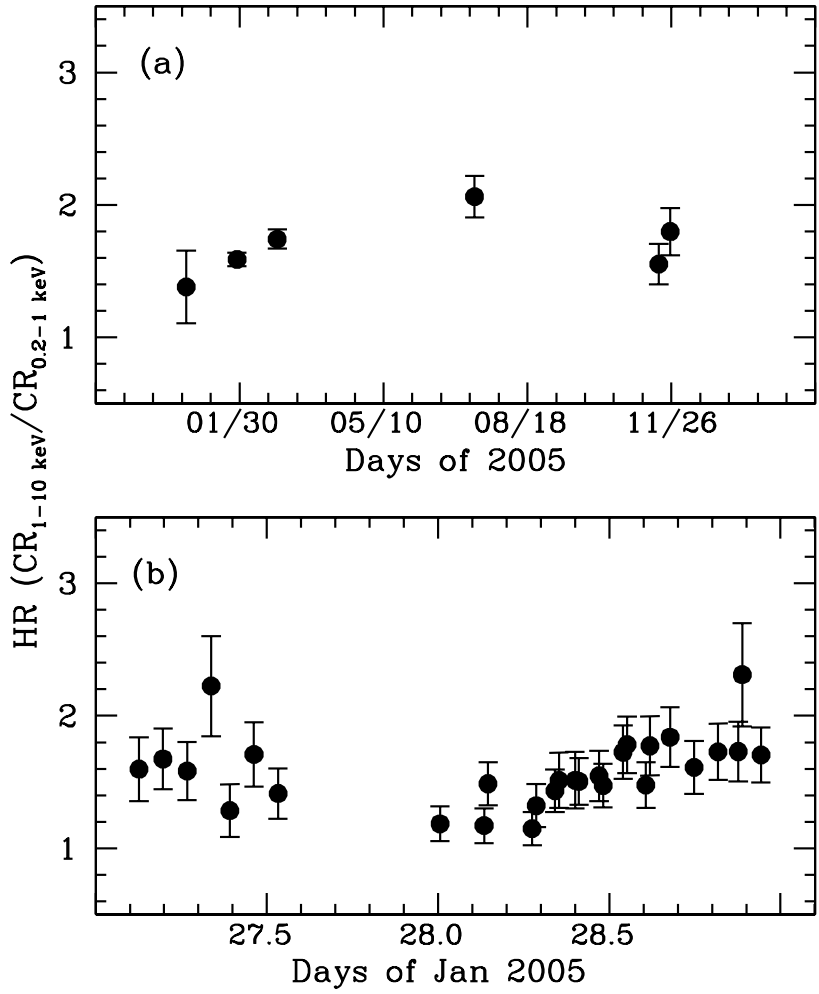}}
\caption{Hardness ratios computed using the 1-10 keV and 0.2-1 keV
count rates for  (a) the whole monitoring;  (b) the observation
of January 2005.}
\end{figure*}


\begin{figure*}
\resizebox{\hsize}{!}{\includegraphics[angle=0]{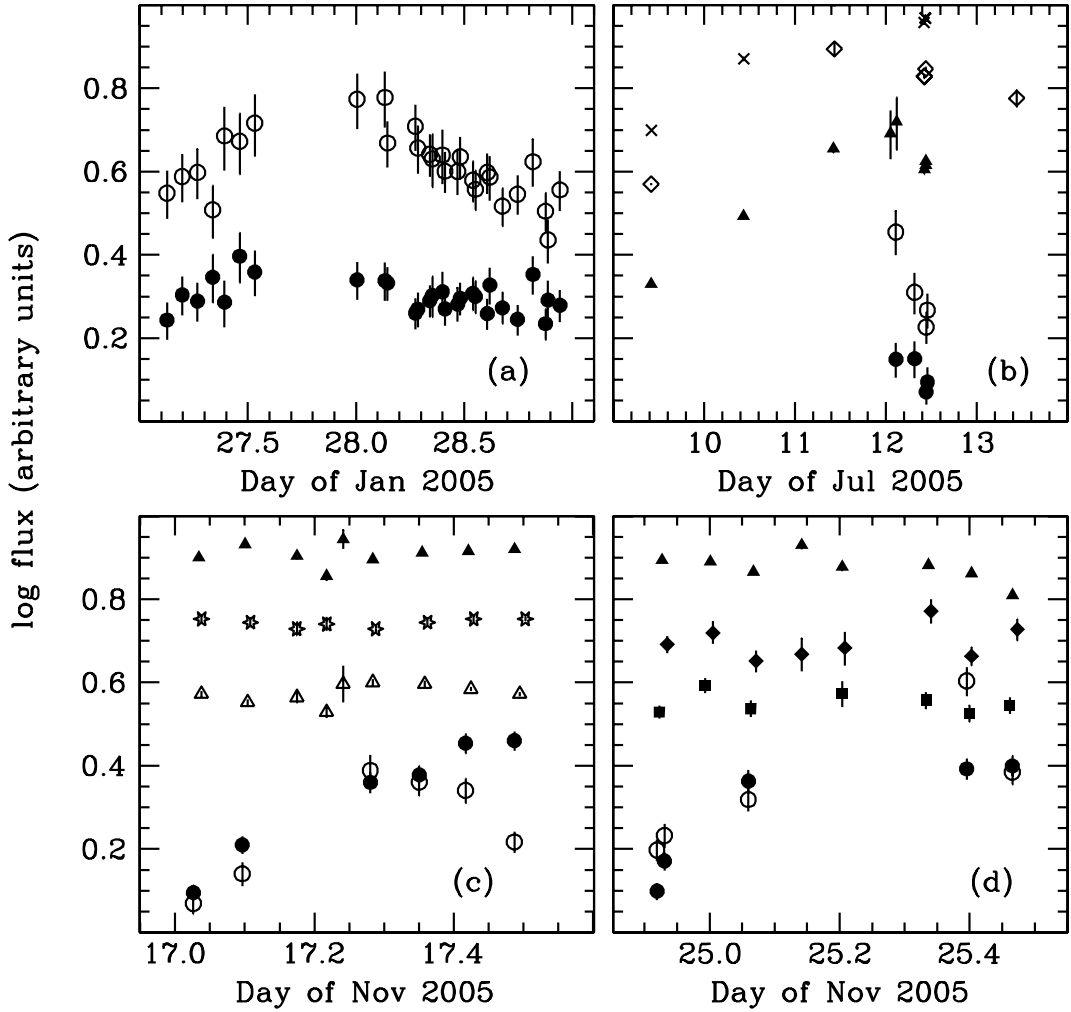}}
\caption{Multiwavelength  light curves  at various  epochs  during the
2005  campaign, in  logarithmic scale,  shifted in  flux  by arbitrary
additive constants: (a) 1-10 keV  (filled circles) and 0.2-1 keV (open
circles) XRT  light curves  in January 2005;  (b) XRT light  curves in
July 2005  (symbols as  in panel (a)),  and simultaneous REM  and UVOT
V-band (filled triangles), REM R- (open diamonds) and I-band (crosses)
light  curves.   Note  the   optical  increase  preceding   the  X-ray
observation and the correlated X-ray  and optical decay; (c) XRT light
curves on  17 November  2005 (symbols  as in panel  (a)), and  UVOT U-
(open  triangles),  B- (stars)  and  V-band  (filled triangles)  light
curves; (d)  XRT light  curves in 24-25  November 2005 (symbols  as in
panel  (a)),  and  UVOT light  curves  in  the  W2 (1930  \AA,  filled
squares),  W1  (2600  \AA,  filled  diamonds) and  V  filters  (filled
triangles).}
\end{figure*}


\begin{figure*}
\vskip -3.5 true cm
\resizebox{\hsize}{!}{\includegraphics{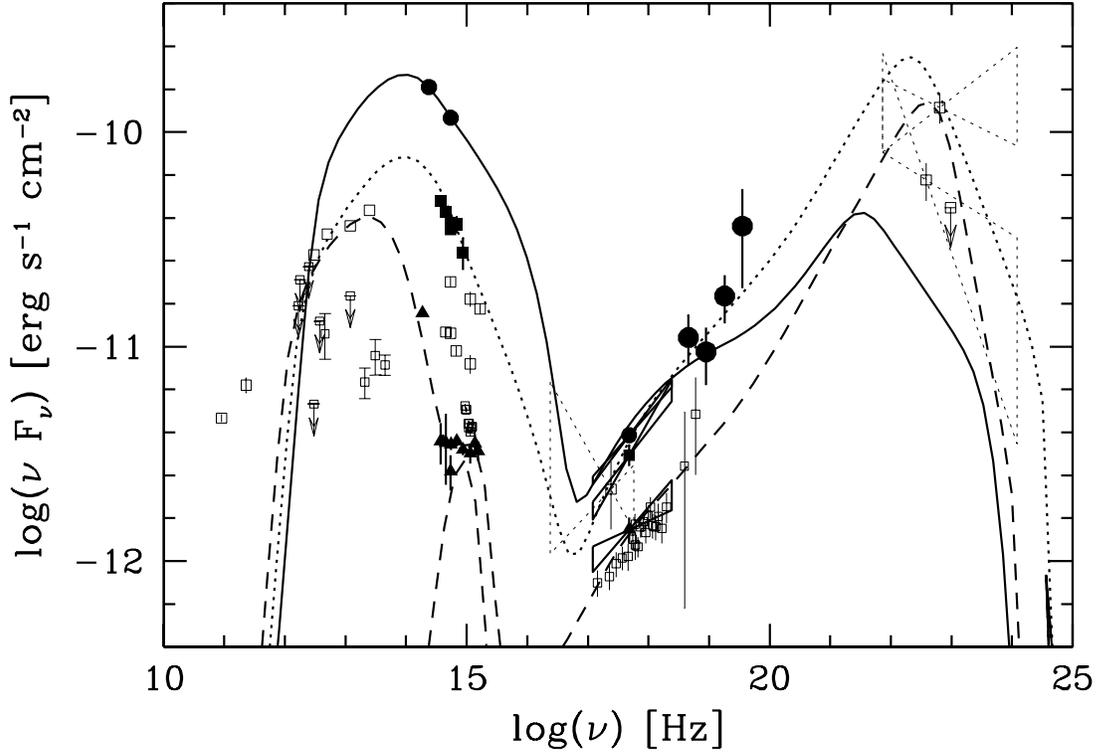}}
\vskip -1 true cm
\caption{Observed  spectral energy  distributions of  \pks\ on  24-25 February
2005 (small filled circles),  12 July 2005 
(filled  squares) and 24 November  2005 (filled
triangles).  The big filled circles represent the BAT data. The 
\swift/XRT data are reported along with the 1 $\sigma$ confidence ranges
of their power-law  fits.   Systematic errors of 5\%  and 10\%
have been added in quadrature to the statistical uncertainties associated 
with the UVOT UBV filters and UV filters data points, respectively.
For comparison, 
in lighter,  open squares 
the  multiwavelength data  from  previous epochs are reported (including
CGRO-EGRET spectra), already  discussed in Pian et al.  (2002), and the
non-simultaneous IRAS, ISO, HST-FOS and BeppoSAX data  (Pian et al. 2002; 
Padovani et  al. 2006).  The 1 $\sigma$ 
confidence ranges
of the EGRET spectra are reported as light dashed lines.
The flux uncertainties are  1 $\sigma$ (in some cases they are smaller than the
symbol size).
The  X-ray, UV,
optical and  near-IR data are  corrected for Galactic  extinction (see
text).   The optical  and near-IR  magnitudes have  been  converted to
fluxes following Fukugita, Shimasaku, \& Ichikawa (1995) and 
Bersanelli, Bouchet, \& Falomo (1991),
respectively.  Overplotted are the jet models (Katarzy\'nski \& Ghisellini 2007,
see text) for the energy distributions of 24-25 February 2005
(solid curve), 12 July 2005 (dotted curve), 24 November 2005 (dashed curve). The
thermal component required to account for the observed optical-UV
flux is also reported as a dashed curve.}
\label{sedbat}     
\end{figure*}



\begin{figure*}[!t]
\resizebox{\hsize}{!}{\includegraphics{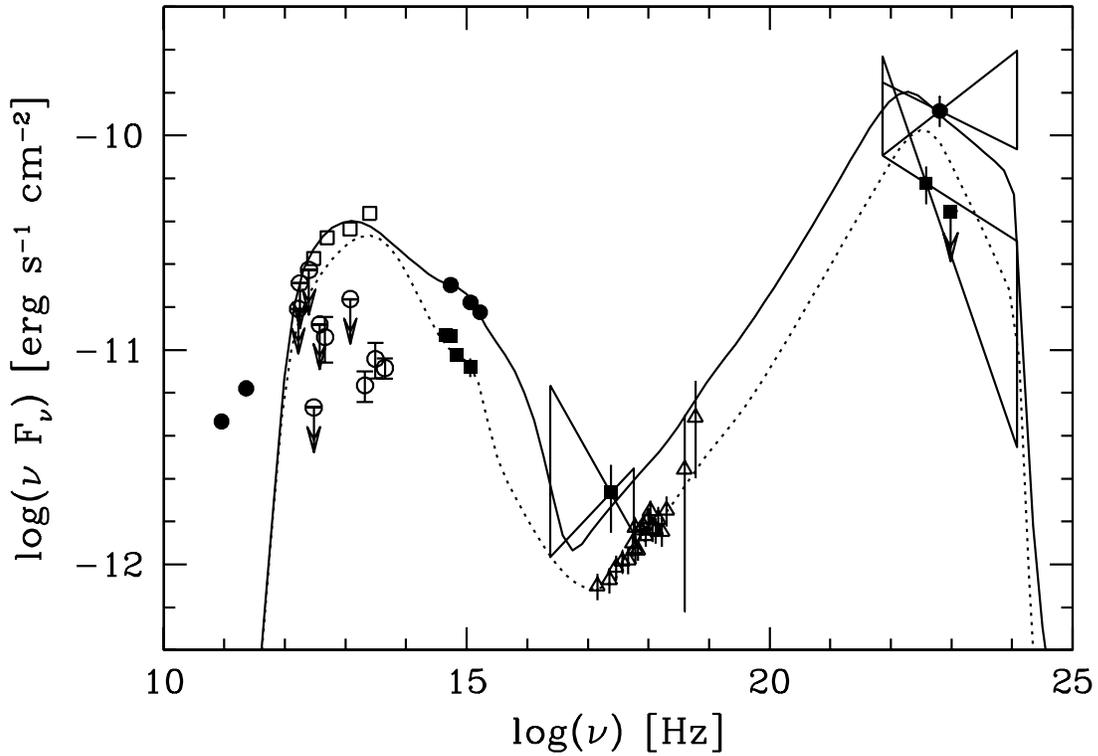}}
\caption{Historical spectral energy  distributions of  \pks.  The data are
the same as those presented in Fig.~5 as open squares. Here we distinguish
them according to the observation epoch: the gamma-ray 
(CGRO-EGRET), soft X-ray (ROSAT), UV (IUE),
optical and millimetric data have been taken nearly simultaneously in 1991-1992
(filled squares) and 1995 (filled circles).  The far-infrared data taken
by IRAS and ISO and the X-ray BeppoSAX data are not simultaneous and are
represented as open squares, open circles and open triangles, respectively (see
Pian et al. 2002, and references therein; Padovani et al. 2006). 
As in Fig.~5, the data have been 
modelled according to Katarzy\'nski \& Ghisellini (2007). The model curves for
the 1991-1992 and 1995 states are dotted and solid, respectively.}
\label{sedold}   
\end{figure*}
\clearpage

 
\begin{figure*}[!t]
\resizebox{\hsize}{!}{\includegraphics{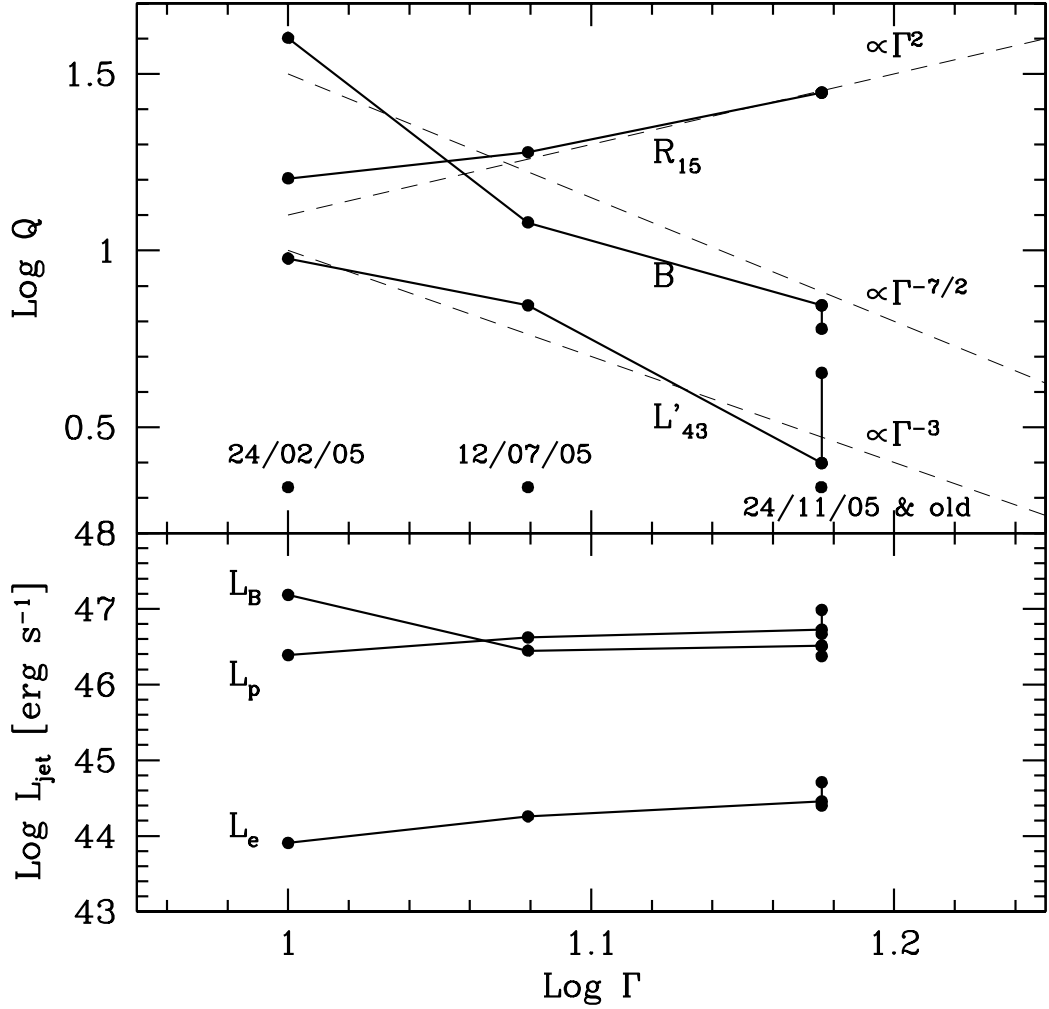}}
\vskip -0.5 true cm
\caption{
{\it Top panel:}  The logarithms of 3 quantities ("Q") are reported as 
a function of the logarithm of the bulk Lorentz factor:
the size of the emitting source $R_{15}$ in units of $10^{15}$ cm, the value of the 
magnetic field $B$ in Gauss, and the injected power 
$L^\prime_{43}$ 
(in the comoving frame) in the 
form of relativistic particles, in units of  $10^{43}$  erg~s$^{-1}$,  as used for our modelling.
The dashed lines represent  the relationships predicted by the Katarzy\'nski \& Ghisellini (2007) model.
The labelled dates identify the specific model/state of the source
(see  Table~3). 
{\it Bottom panel:}
The power carried by the jet in the form of magnetic field 
($L_B$), cold protons ($L_{\rm p}$), relativistic electrons ($L_{\rm e}$)
resulting from our modelling, as a function of the bulk Lorentz factor.
}
\label{sequence}        
\end{figure*}
\vfill


\begin{deluxetable}{ccccccc}
\rotate
\tablenum{1}
\tablewidth{0pt}
\tablecaption{\emph{Swift}/XRT Observation log\tablenotemark{a}}
\tablehead{
\colhead{Start time (UT)}             & 
\colhead{End time (UT)}               &
\colhead{Exposure}                    & 
\colhead{Mean Flux\tablenotemark{b}}  &
\colhead{Counts}                      & 
\colhead{Spectral index}              & 
\colhead{$\chi^{2}_{\rm red}$ (d.o.f.)} \\
\colhead{(yyyy-mm-dd hh:mm:ss)}       & 
\colhead{(yyyy-mm-dd hh:mm:ss)}       & 
\colhead{(s)}                         & 
\colhead{(0.5--10 keV)}               & 
\colhead{(0.2--10 keV)}               & 
\colhead{}                            & 
\colhead{C-stat (\%)}}
\startdata
2004-12-23 23:37:34 & 2004-12-23 23:59:40 & 1029  & $0.67_{-0.50}^{+0.83}$ &
168  & $0.75\pm0.18$ & 372.7 (32.2)\tablenotemark{c} \\
2005-01-27 00:05:19 & 2005-01-27 22:37:41 & 7248  & $1.49_{-0.08}^{+0.10}$ &
2660 & $0.73\pm0.05$ & 1.06 (101)   \\
2005-01-28 00:08:16 & 2005-01-28 23:03:41 & 22663 & $1.38\pm0.05$          &
7794 & $0.76\pm0.03$ & 1.12 (244)  \\
2005-02-24 16:19:51 & 2005-02-25 12:00:40 & 18592 & $1.26\pm0.05$          &
5420 & $0.65\pm0.03$ & 0.99 (198)  \\
2005-07-12 01:13:01 & 2005-07-12 11:12:59 & 4977  & $1.05\pm0.11$          &
1003  & $0.57\pm0.07$ & 0.93 (44)     \\
2005-11-17 00:45:12 & 2005-11-17 12:11:11 & 6373  & $0.39\pm0.06$          &
556  & $0.75^{+0.11}_{-0.10}$ & 1.16 (23) \\
2005-11-24 22:14:00 & 2005-11-25 11:25:57 & 6243  & $0.44_{-0.06}^{+0.07}$ &
583  & $0.77\pm0.10$ & 1.00 (24)     
\enddata
\tablenotetext{a}{All observations were carried out with XRT in PC observing mode.}
\tablenotetext{b}{Unabsorbed flux in units of $10^{-11}$
erg cm$^{-2}$ s$^{-1}$. The effects of PSF-loss and vignetting  were taken
into account in  the count rate to flux conversion.}
\tablenotetext{c}{Cash statistic (C-stat) and percentage
of Monte Carlo realizations that had statistic $<$ C-stat,
for this entry only.}
\end{deluxetable}

\clearpage

\begin{deluxetable}{rcc}
\tablenum{2}
\tablewidth{0pt}
\tablecaption{\emph{Swift}/UVOT and REM observations of PKS~0537-441 in July and November 2005\tablenotemark{a}}
\tablehead{
\colhead{Date (UT)}                   & 
\colhead{MJD\tablenotemark{b}}        & 
\colhead{magnitude\tablenotemark{c}}}
\startdata
\cutinhead{UVOT UVW2 filter (1930 \AA)}
2005 Nov 24.9243 &  53698.9243  &    $17.02 \pm      0.04$\tablenotemark{d} \\

         24.9958 &  53698.9958  &    $16.86 \pm      0.04$ \\
         25.0646 &  53699.0646  &    $17.00 \pm      0.05$ \\
         25.2042 &  53699.2042  &    $16.91 \pm      0.07$ \\
         25.3313 &  53699.3313  &    $16.95 \pm      0.05$ \\
         25.3979 &  53699.3979  &    $17.03 \pm      0.05$ \\
         25.4653 &  53699.4653  &    $16.98 \pm      0.05$ \\
\cutinhead{UVOT UVM2 filter (2200 \AA)}
2005 Nov 24.9312 & 53698.9312  &    $16.77 \pm      0.05$ \\
         25.0007 & 53699.0007  &    $16.88 \pm      0.06$ \\
         25.0688 & 53699.0688  &    $16.64 \pm      0.06$ \\
         25.1382 & 53699.1382  &    $16.60 \pm      0.08$ \\
         25.2063 & 53699.2063  &    $16.59 \pm      0.09$ \\
         25.3361 & 53699.3361  &    $16.89 \pm      0.07$ \\
         25.4028 & 53699.4028  &    $16.76 \pm      0.06$ \\
         25.4701 & 53699.4701  &    $16.73 \pm      0.06$ \\
\cutinhead{UVOT UVW1 filter (2600 \AA)}
2005 Nov 24.9354 &  53698.9354  &    $16.60 \pm      0.05$ \\
         25.0035 &  53699.0035  &    $16.53 \pm      0.06$ \\
         25.0715 &  53699.0715  &    $16.70 \pm      0.07$ \\
         25.1396 &  53699.1396  &    $16.66 \pm      0.10$ \\
         25.2076 &  53699.2076  &    $16.62 \pm      0.10$ \\
         25.3389 &  53699.3389  &    $16.40 \pm      0.06$ \\
         25.4056 &  53699.4056  &    $16.67 \pm      0.06$ \\
         25.4729 &  53699.4729  &    $16.51 \pm      0.06$ \\
\cutinhead{UVOT U filter (3450 \AA)}
2005 Jul 12.0530 &  53563.0530  &    $14.05   \pm    0.16$ \\
         12.1190 &  53563.1190  &    $14.03   \pm    0.16$ \\ 
     Nov 17.0368 &  53691.0368  &    $16.22   \pm    0.03$ \\
         17.1042 &  53691.1042  &    $16.27   \pm    0.03$ \\
         17.1736 &  53691.1736  &    $16.24   \pm    0.05$ \\
         17.2201 &  53691.2201  &    $16.33   \pm    0.06$ \\
         17.2424 &  53691.2424  &    $16.16   \pm    0.16$ \\
         17.2896 &  53691.2896  &    $16.15   \pm    0.03$ \\
         17.3590 &  53691.3590  &    $16.16   \pm    0.03$ \\
         17.4264 &  53691.4264  &    $16.19   \pm    0.02$ \\
         17.4986 &  53691.4986  &    $16.22   \pm    0.02$ \\
         24.9375 &  53698.9375  &    $16.49   \pm    0.05$ \\
         25.0056 &  53699.0056  &    $16.56   \pm    0.06$ \\
         25.0736 &  53699.0736  &    $16.37   \pm    0.06$ \\
         25.1409 &  53699.1409  &    $16.50   \pm    0.09$ \\
         25.2083 &  53699.2083  &    $16.62   \pm    0.10$ \\
         25.3409 &  53699.3409  &    $16.57   \pm    0.06$ \\
         25.4076 &  53699.4076  &    $16.60   \pm    0.06$ \\
         25.4743 &  53699.4743  &    $16.63   \pm    0.06$ \\
\cutinhead{UVOT B filter (4350 \AA)}
2005 Jul 12.0560 &  53563.0560   &  $14.81   \pm 0.08$ \\
         12.1220 &  53563.1220   &  $14.76   \pm 0.08$ \\
     Nov 17.0403 &  53691.0403   &  $16.97   \pm 0.03$ \\
         17.1076 &  53691.1076   &  $16.99   \pm 0.04$ \\
         17.1750 &  53691.1750   &  $17.03   \pm 0.07$ \\
         17.2208 &  53691.2208   &  $17.00   \pm 0.07$ \\
         17.2917 &  53691.2917   &  $17.03   \pm 0.04$ \\
         17.3625 &  53691.3625   &  $16.99   \pm 0.03$ \\
         17.4306 &  53691.4306   &  $16.97   \pm 0.03$ \\
         17.5056 &  53691.5056   &  $16.97   \pm 0.02$ \\
         24.9389 &  53698.9389   &  $17.36   \pm 0.06$ \\
         25.0069 &  53699.0069   &  $17.31   \pm 0.08$ \\
         25.0743 &  53699.0743   &  $17.37   \pm 0.10$ \\
         25.1409 &  53699.1409   &  $17.25   \pm 0.14$ \\
         25.3417 &  53699.3417   &  $17.34   \pm 0.08$ \\
         25.4090 &  53699.4090   &  $17.31   \pm 0.08$ \\
         25.4757 &  53699.4757   &  $17.22   \pm 0.08$ \\
\cutinhead{UVOT V filter (5460 \AA)}
2005 Jul 12.0500 &  53563.0500   &  $14.32  \pm     0.09$ \\
         12.1150 &  53563.1150   &  $14.25  \pm     0.09$ \\
     Nov 17.0340 &  53691.0340   &  $16.47  \pm     0.04$ \\
         17.1014 &  53691.1014   &  $16.39  \pm     0.04$  \\
         17.1729 &  53691.1729   &  $16.46  \pm     0.07$  \\
         17.2194 &  53691.2194   &  $16.58  \pm     0.15$  \\
         17.2424 &  53691.2424   &  $16.36  \pm     0.21$  \\
         17.2875 &  53691.2875   &  $16.48  \pm     0.05$  \\
         17.3556 &  53691.3556   &  $16.44  \pm     0.04$  \\
         17.4222 &  53691.4222   &  $16.43  \pm     0.04$  \\
         17.4910 &  53691.4910   &  $16.42  \pm     0.03$  \\
         24.9286 &  53698.9286   &  $16.81  \pm     0.04$  \\
         25.0000 &  53699.0000   &  $16.82  \pm     0.09$  \\
         25.0681 &  53699.0681   &  $16.88  \pm     0.10$  \\
         25.1382 &  53699.1382   &  $16.72  \pm     0.13$  \\
         25.2055 &  53699.2055   &  $16.85  \pm     0.15$  \\
         25.3354 &  53699.3354   &  $16.84  \pm     0.10$ \\
         25.4021 &  53699.4021   &  $16.89  \pm     0.10$ \\
         25.4694 &  53699.4694   &  $17.02  \pm     0.11$ \\
\cutinhead{REM V filter (5505 \AA)}
2005 Jul 07.4375 &  53558.4375   &   $15.72  \pm     0.03$ \\
         09.4219 &  53560.4219   &   $15.21  \pm     0.02$ \\
         10.4336 &  53561.4336   &   $14.80  \pm     0.02$ \\
         11.4297 &  53562.4297   &   $14.40  \pm     0.02$ \\
         12.4297 &  53563.4297   &   $14.52  \pm     0.02$ \\
         12.4375 &  53563.4375   &   $14.47  \pm     0.02$ \\
         12.4414 &  53563.4414   &   $14.50  \pm     0.02$ \\
         22.3984 &  53573.3984   &   $15.47  \pm     0.04$ \\
         22.4336 &  53573.4336   &   $15.73  \pm     0.03$ \\
     Nov 07.2494 &  53681.2494   &   $16.93  \pm     0.11$ \\
         19.2346 &  53693.2346   &   $17.07  \pm     0.13$ \\
         20.2174 &  53694.2174   &   $17.17  \pm     0.14$ \\
         30.2021 &  53704.2021   &   $17.46  \pm     0.17$ \\
\cutinhead{REM R filter (6588 \AA)}
2005 Jul 07.4375 &  53558.4375  & $15.06  \pm  0.01$ \\
         09.4219 &  53560.4219  & $14.56  \pm  0.01$ \\
         11.4336 &  53562.4336  & $13.75  \pm  0.04$ \\
         12.4258 &  53563.4258  & $13.91  \pm  0.01$ \\
         12.4297 &  53563.4297  & $13.91  \pm  0.01$ \\
         12.4375 &  53563.4375  & $13.87  \pm  0.01$ \\
         13.4414 &  53564.4414  & $14.04  \pm  0.06$ \\
         22.4023 &  53573.4023  & $14.87  \pm  0.02$ \\
     Nov 06.2533 &  53680.2533  & $16.39  \pm  0.07$ \\
         07.2522 &  53681.2522  & $16.39  \pm  0.04$ \\
         18.2249 &  53692.2249  & $16.41  \pm  0.06$ \\
         19.2452 &  53693.2452  & $16.20  \pm  0.06$ \\
         20.2280 &  53694.2280  & $16.35  \pm  0.07$ \\
         30.2127 &  53704.2127  & $16.92  \pm  0.08$ \\
\cutinhead{REM I filter (8060 \AA)}
2005 Jul 07.4375 &  53558.4375 &  $14.40  \pm  0.01$ \\
         09.4219 &  53560.4219 &  $13.91  \pm  0.01$ \\
         10.4375 &  53561.4375 &  $13.48  \pm  0.01$ \\
         11.4336 &  53562.4336 &  $13.14  \pm  0.03$ \\
         12.4258 &  53563.4258 &  $13.26  \pm  0.01$ \\
         12.4336 &  53563.4336 &  $13.23  \pm  0.01$ \\
         12.4414 &  53563.4414 &  $13.24  \pm  0.01$ \\
         20.4414 &  53571.4414 &  $13.82  \pm  0.03$ \\
         22.4062 &  53573.4062 &  $14.13  \pm  0.02$ \\
         22.4375 &  53573.4375 &  $14.22  \pm  0.02$ \\
     Nov 06.2559 &  53680.2559 &  $15.60  \pm  0.08$ \\
         07.2549 &  53681.2549 &  $15.78  \pm  0.07$ \\
         18.2355 &  53692.2355 &  $15.82  \pm  0.08$ \\
         19.2560 &  53693.2560 &  $15.72  \pm  0.07$ \\
         20.2386 &  53694.2386 &  $15.89  \pm  0.10$ \\
         30.2235 &  53704.2235 &  $16.23  \pm  0.09$ \\
\cutinhead{REM H filter (16000 \AA)}
2005 Nov 02.1718 &  53676.1718 &  $12.77  \pm  0.07$ \\
         02.1738 &  53676.1738 &  $12.89  \pm  0.08$ \\
         02.1759 &  53676.1759 &  $12.93  \pm  0.08$ \\
         02.1805 &  53676.1805 &  $12.89  \pm  0.10$ \\
         02.1820 &  53676.1820 &  $12.70  \pm  0.06$ \\
         02.1917 &  53676.1917 &  $12.80  \pm  0.07$ \\
         02.1941 &  53676.1941 &  $12.75  \pm  0.07$ \\
         02.1962 &  53676.1962 &  $12.85  \pm  0.07$ \\
         02.2003 &  53676.2003 &  $12.77  \pm  0.08$ \\
         02.2895 &  53676.2895 &  $12.75  \pm  0.07$ \\
         02.2957 &  53676.2957 &  $13.05  \pm  0.06$ \\
         02.3027 &  53676.3027 &  $12.88  \pm  0.08$ \\
         02.3048 &  53676.3048 &  $12.78  \pm  0.07$ \\
         02.3068 &  53676.3068 &  $12.73  \pm  0.06$ \\
         02.3089 &  53676.3089 &  $12.96  \pm  0.08$ \\
         03.2822 &  53677.2822 &  $12.95  \pm  0.07$ \\
         03.2864 &  53677.2864 &  $12.78  \pm  0.06$ \\
         03.2885 &  53677.2885 &  $12.83  \pm  0.07$ \\
         03.2905 &  53677.2905 &  $12.72  \pm  0.06$ \\
         03.2929 &  53677.2929 &  $12.66  \pm  0.06$ \\
         03.2950 &  53677.2950 &  $12.91  \pm  0.09$ \\
         03.3006 &  53677.3006 &  $12.78  \pm  0.07$ \\
         03.3027 &  53677.3027 &  $12.70  \pm  0.08$ \\ 
         03.3047 &  53677.3047 &  $12.63  \pm  0.07$ \\
         03.3068 &  53677.3068 &  $12.80  \pm  0.07$ \\
         03.3088 &  53677.3088 &  $12.67  \pm  0.06$ \\
         03.3112 &  53677.3112 &  $12.90  \pm  0.08$ \\
         03.3133 &  53677.3133 &  $12.98  \pm  0.07$ \\
         03.3154 &  53677.3154 &  $12.84  \pm  0.07$ \\
         03.3174 &  53677.3174 &  $12.98  \pm  0.07$ \\
         03.3195 &  53677.3195 &  $12.95  \pm  0.07$ \\
         03.3219 &  53677.3219 &  $13.08  \pm  0.07$ \\
         03.3240 &  53677.3240 &  $12.75  \pm  0.07$ \\
         03.3260 &  53677.3260 &  $12.79  \pm  0.06$ \\
         03.3281 &  53677.3281 &  $12.74  \pm  0.06$ \\
         03.3302 &  53677.3302 &  $12.84  \pm  0.08$ \\
         17.1083 &  53691.1083 &  $12.42  \pm  0.08$ \\
         18.2154 &  53692.2154 &  $12.68  \pm  0.06$ \\
         18.2208 &  53692.2208 &  $12.64  \pm  0.07$ \\
         18.2229 &  53692.2229 &  $12.61  \pm  0.08$ \\
         18.2249 &  53692.2249 &  $12.70  \pm  0.07$ \\
2005 Nov 18.2273 &  53692.2273 &  $12.64  \pm  0.06$ \\
         18.2296 &  53692.2296 &  $12.63  \pm  0.07$ \\
         18.2314 &  53692.2314 &  $12.74  \pm  0.06$ \\
         18.2335 &  53692.2335 &  $12.73  \pm  0.07$ \\
         18.2355 &  53692.2355 &  $12.61  \pm  0.06$ \\
         18.2379 &  53692.2379 &  $12.71  \pm  0.07$ \\
         18.2399 &  53692.2399 &  $12.51  \pm  0.06$ \\
         18.2420 &  53692.2420 &  $12.61  \pm  0.07$ \\
         18.2440 &  53692.2440 &  $12.70  \pm  0.06$ \\
         18.2461 &  53692.2461 &  $12.60  \pm  0.07$ \\
         18.2626 &  53692.2626 &  $12.68  \pm  0.07$ \\
         18.2654 &  53692.2654 &  $12.68  \pm  0.07$ \\
         18.2688 &  53692.2688 &  $12.62  \pm  0.07$ \\
         19.2350 &  53693.2350 &  $12.65  \pm  0.07$ \\
         19.2371 &  53693.2371 &  $12.54  \pm  0.08$ \\
         19.2391 &  53693.2391 &  $12.72  \pm  0.08$ \\
         19.2412 &  53693.2412 &  $12.65  \pm  0.08$ \\
         19.2432 &  53693.2432 &  $12.55  \pm  0.07$ \\
         19.2453 &  53693.2453 &  $12.59  \pm  0.07$ \\
         19.2476 &  53693.2476 &  $12.52  \pm  0.07$ \\
         19.2497 &  53693.2497 &  $12.54  \pm  0.08$ \\
         19.2518 &  53693.2518 &  $12.66  \pm  0.07$ \\
         19.2605 &  53693.2605 &  $12.54  \pm  0.06$ \\
         20.2220 &  53694.2220 &  $12.74  \pm  0.07$ \\
         20.2240 &  53694.2240 &  $12.96  \pm  0.07$ \\
         20.2249 &  53694.2249 &  $12.66  \pm  0.06$ \\
         20.2261 &  53694.2261 &  $12.83  \pm  0.07$ \\
         20.2281 &  53694.2281 &  $12.83  \pm  0.06$ \\
         20.2346 &  53694.2346 &  $12.77  \pm  0.06$ \\
         20.2366 &  53694.2366 &  $12.62  \pm  0.05$ \\
         20.2387 &  53694.2387 &  $12.76  \pm  0.06$ \\
         20.2411 &  53694.2411 &  $12.74  \pm  0.06$ \\
         20.2431 &  53694.2431 &  $12.85  \pm  0.07$ \\
         20.2452 &  53694.2452 &  $12.77  \pm  0.07$ \\
         20.2472 &  53694.2472 &  $12.82  \pm  0.07$ \\
         20.2493 &  53694.2493 &  $12.66  \pm  0.06$ \\
         20.2513 &  53694.2513 &  $12.74  \pm  0.07$ \\
         30.1977 &  53704.1977 &  $13.10  \pm  0.08$ \\
         30.2115 &  53704.2115 &  $13.20  \pm  0.10$ \\
         30.2135 &  53704.2135 &  $13.33  \pm  0.11$ \\
         30.2179 &  53704.2179 &  $13.37  \pm  0.10$ \\
         30.2200 &  53704.2200 &  $13.08  \pm  0.10$ \\
         30.2371 &  53704.2371 &  $13.20  \pm  0.10$ 
\enddata
\tablenotetext{a}{Typical exposure times are 4-5 minutes, both for UVOT and REM.}
\tablenotetext{b}{Modified Julian Date.  We used the convention MJD = JD - 2,400,000.5.}
\tablenotetext{c}{Not corrected for Galactic extinction.}
\tablenotetext{d}{Errors represent 1 $\sigma$ statistical uncertainties. }\\
\end{deluxetable}


\begin{deluxetable}{lllllll}
\tablenum{3}
\tablewidth{0pt}
\tablecaption{Model parameters}
\tablehead{
\colhead{Feb 05}                   & 
\colhead{Jul 05}                   & 
\colhead{Nov 05}                   & 
\colhead{1991--1992}               & 
\colhead{1995}                     & 
\colhead{Units}} 
\startdata
$\Gamma$             &10    &12   &15    &15    &15    &       \\
$R$                  &16    &19   &28    &28    &28    & [$10^{15}$ cm]   \\
$B$                  &40    &12   &7     &6     &7     & [G]    \\
$\gamma_{\rm b}$     &250   &500  &380   &200   &310   &     \\
$\gamma_{\rm max}$   &3e3   &6e3  &2e4   &7e3   &2e4   &     \\
$p$                  &3.8   &4.4  &6.2   &3.5   &4.2   &     \\
$L^\prime_{\rm inj}$ &0.095 &0.07 &0.025 &0.045 &0.03  & [$10^{45}$ erg s$^{-1}$] \\
$\nu_{\rm ext}$      &0.3   &0.5  &1.5   &1.5   &1.5   & [$10^{15}$ Hz]\\
$U_{\rm ext}$        &0.085 &0.069&0.016 &0.016 &0.016 & [erg cm$^{-3}$] \\
$\delta$             &15.7 &17.2  &18.6  &18.6  &18.6  &              \\
$L_{B}$              &253  &28    &32.3  &23.8  &32.3  &[$10^{45}$ erg s$^{-1}$]  \\
$L_{\rm p}$          &24.5 &42    &53.2  &96.8  &46.6  &[$10^{45}$ erg s$^{-1}$]  \\
$L_{\rm e}$          &0.08 &0.18  &0.29  &0.51  &0.25  &[$10^{45}$ erg s$^{-1}$]  \\
$L_{\rm rad}$        &8.8  &10.2  &5.6   &9.95  &5.5   &[$10^{45}$ erg s$^{-1}$]   
\enddata
\tablecomments{For all models we have assumed a viewing angle $\theta=3^\circ$ 
and a bolometric luminosity of the accretion disk 
$L_{\rm disk}=1.8\times 10^{46}$ erg s$^{-1}$. 
The energy density of the external radiation and its peak frequency
$\nu_{\rm ext}$ are measured in the observer frame.}

\end{deluxetable}


\newpage

\begin{thebibliography}{}{}

\bibitem[Aharonian et al. 2006]{fah2006} 
Aharonian, F., et al. 2006, A\&A, 455, 461

\bibitem[Albert et al. 2006]{alb2006} 
Albert, J., et al. 2006, ApJ, 648, L105

\bibitem[Bersanelli et al. 1991]{bbf1991} 
Bersanelli, M., Bouchet, P., \& Falomo, E. 1991, A\&A, 252, 854

\bibitem[B{\l}a\.zejowski et al. 2000]{bla2000} 
B{\l}a\.zejowski, M., Sikora, M., Moderski, R., \& Madejski, G. M.
2000, ApJ, 545, 107

\bibitem[B{\l}a\.zejowski et al. 2005]{bla2005} 
B{\l}a\.zejowski, M., et al. 2005, ApJ, 630, 130

\bibitem[Burrows et al.(2005)]{XRT}
Burrows, D.~N., et al.\ 2005, 
Space Sci.\ Rev., 120, 165

\bibitem[Cardelli, Clayton, \& Mathis (1989)]{jc1989} 
Cardelli, J. A., Clayton, G. C., \& Mathis, J. S. 1989, ApJ, 345, 245 

\bibitem[Cash(1979)]{Cash79}
Cash, W.\ 1979, \apj, 228, 939 

\bibitem[Celotti et al. 2007]{cgf07} 
Celotti, A., Ghisellini, G., \& Fabian, A.C. 2007, MNRAS, in press 
(astro--ph/0611439)

\bibitem[Chincarini et al. 2003]{chi03} 
Chincarini, G., et al. 2003, 
The ESO Messenger, No.113, p. 40-44

\bibitem[Covino et al. 2004]{cov04} 
Covino, S., et al. 2004, AN, 325, 543

\bibitem[Covino et al. 2005]{cov05} 
Covino, S., et al. 2005, The Astronomer's Telegram N. 564

\bibitem[Dermer \& Schlickeiser 1993]{ds93} 
Dermer, C.D., \& Schlickeiser, R. 1993, ApJ, 416, 453

\bibitem[Dermer \& Atoyan 2004]{cddaa04} 
Dermer, C.D., \& Atoyan, A. 2004, ApJ, 611, L9 

\bibitem[Dolcini et al. 2005]{dolc05} 
Dolcini, A., et al. 2005, A\&A, 443, L33

\bibitem[Fuhrmann et al. 2006]{fuh06} 
Fuhrmann, L., et al. 2006, A\&A, 445, L1

\bibitem[Fukugita et al. 1995]{fuk95} 
Fukugita, M., Shimasaku, K., \& Ichikawa, T. 1995, PASP, 107, 945

\bibitem[Gehrels et al. 2004]{gehrels04} 
Gehrels, N., et al. 2004, ApJ, 611, 1005

\bibitem[Ghisellini 1989]{gg89} 
Ghisellini, G. 1989, MNRAS, 238, 449

\bibitem[Ghisellini \& Madau 1996]{ggpm96} 
Ghisellini, G., \& Madau, P. 1996, MNRAS, 280, 67    

\bibitem[Ghisellini  et al. 1998]{gg+98} 
Ghisellini, G. 1998, Celotti, A., Fossati, G. Maraschi, L., \& Comastri, A. 1998, MNRAS, 301, 451 

\bibitem[Ghisellini 1999]{gg99} 
Ghisellini, G. 1999, AN, 320, 232

\bibitem[Ghisellini et al. 2002]{gg+02} 
Ghisellini, G., Celotti, A., \& Costamante, L. 2002, A\&A,  386,  833

\bibitem[Giommi et al. 2006]{giommi06} 
Giommi, P., et al. 2006, A\&A, 456, 911

\bibitem[Guetta et al. 2004]{gglc04} 
Guetta, D., Ghisellini, G., Lazzati, D. \& Celotti, A. 2004, A\&A, 421, 877

\bibitem[Hartman et al. 1999]{hbb99} 
Hartman, R.C., et al. 1999, ApJS, 123, 79

\bibitem[Hill et al.(2004)]{xrtmodes} 
Hill, J.~E., et al.\ 2004, \procspie, 5165, 217 

\bibitem[Katarzy\'nski \& Ghisellini (2007)]{katghi06} 
Katarzy\'nski, K., \& Ghisellini, G. 2007, A\&A, 463, 529

\bibitem[Kato et al. (2006)]{kato06} 
Kato, T., Kusunose, M., Takahara, F. 2006, ApJ, 638, 653

\bibitem[Krawczynski et al. (2004)]{kraw04} 
Krawczynski, H., et al. 2004, ApJ, 601, 151

\bibitem[Massaro et al. (2006)]{mas06} 
Massaro, E.,  Tramacere, A., Perri, M., Giommi, P., \& Tosti, G. 
2006, A\&A, 448, 861

\bibitem[Moretti et al.(2006)]{centroids}
Moretti, A., et al.\ 2006, A\&A, 448, L9 

\bibitem[Murphy et al.(1996)]{murphy96}
Murphy, E.M., Lockman, F.J., Laor, A., \& Elvis, M. 1996, ApJS, 105, 369


\bibitem[Padovani et al. (2006)]{pp+06} 
Padovani, P., Giommi, P., \'Abrah\'am, P., Csizmadia, S., \& Mo\'or, A. 
2006, A\&A, 456, 131

\bibitem[Pian et al. (1998)]{pian98} 
Pian, E., et al. 1998, ApJ, 492, L17

\bibitem[Pian et al. (1999)]{pum99} 
Pian, E., et al. 1999, ApJ, 521, 112 

\bibitem[Pian et al. (2002)]{ep02} 
Pian, E., et al. 2002, A\&A, 392, 407

\bibitem[Pian et al. (2005)]{ep05} 
Pian, E., Falomo, R., \& Treves, A. 2005, MNRAS, 361, 919

\bibitem[Pian et al. (2006)]{pian+06} 
Pian, E., et al. 2006, A\&A, 449, L21

\bibitem[Raiteri et al. (2006)]{rait+06} 
Raiteri, C.M., et al. 2006, A\&A, 459, 731

\bibitem[Roming et~al.(2005)]{UVOT}
{Roming}, P.~W.~A., {et~al.} 2005, Space Sci. Rev., 120, 95

\bibitem[Sambruna et al. 2006]{rms06} 
Sambruna, R.M., et al. 2006, ApJ, 646, 23

\bibitem[Schlegel et al. 1998]{sfd1998} 
Schlegel, D.J., Finkbeiner, D.P., \& Davis, M. 1998, ApJ, 500, 525

\bibitem[Sikora et al. 1994]{sbr94} 
Sikora, M., Begelman, M.C., \& Rees, M.J. 1994, ApJ, 421, 153

\bibitem[Sokolov \& Marscher 2005]{soma05} 
Sokolov, A., \& Marscher, A.P. 2005, ApJ, 629, 52

\bibitem[Spada et al. 2001]{spada+01} 
Spada, M., Ghisellini, G., Lazzati, D. \& Celotti, A. 2001, MNRAS, 325, 1559

\bibitem[Tagliaferri et al. 2003]{tag03} 
Tagliaferri, G., et al. 2003, A\&A, 400, 477

\bibitem[Tramacere et al. 2006]{tra06} 
Tramacere, A., et al. 2006, A\&A, in press (astro-ph/0611276)

\bibitem[Treves et al. 1993]{at93} 
Treves, A., et al.  1993, ApJ, 406, 447

\bibitem[Ulrich et al. 1997]{umu97} 
Ulrich, M.-H., Maraschi, L., \& Urry, C.M. 1997, ARA\&A, 35, 445

\bibitem[Zerbi et al. 2001]{zer2001} 
Zerbi, F.M., et al. 2001, AN, 322, 275

\end{thebibliography}
\end{document}